\documentclass[11pt, oneside]{article} 
\usepackage{geometry} 
\usepackage{graphicx} 

\usepackage{amssymb}

\title{\huge G3Ms: Generalized Mean Market Makers}
\author{Daniel Z. Zanger\\ MarketX LLC\\danielzanger@gmail.com}
\date{}
\begin{document}
\maketitle
\noindent \bf Abstract:\rm\,In the Decentralized Finance (DeFi) setting, we present a new parametrized family of Constant Function Market Makers (CFMMs) which we call 
the Generalized Mean Market Makers (G3Ms), based on the generalized means. The G3Ms are intermediate between the Arithmetic Mean and Geometric Mean
CFMM models, which G3Ms incorporate as special cases. We also present an extension of the G3Ms, based on the so-called Generalized f-Means, called Generalized f-Mean Market Makers (Gf3Ms). We show in addition that the G3Ms possess certain properties preferable to those exhibited by either the Arithmetic Mean CFMM or the Geometric Mean CFMM alone.

\section{Introduction}

Gaining wider use and broader popularity in recent years, Decentralized Finance (DeFi) involves innovative, emerging financial technologies intended 
to facilitate a wide range of peer-to-peer transactions. DeFi critically relies in its design on blockchain and related distributed ledger technologies, 
and one of DeFi's core application domains is the Decentralized Exchange (DEX) (see [9]). Indeed, a principal use case for the DEX platform concept
is as a medium for buying and selling cryptocurrencies in which market participants do not require a trusted third party to execute transactions. 

Among the the most widely-applied types of DEX architectures are those employing Automated Market Maker (AMM) protocols [9].
AMMs utilize liquidity pools instead of a traditional market of buyers and sellers to enable trading of digital assets
without intermediaries. The operation of an AMM relies on a trading function, 
the nature of which governs the trading dynamics of the exchange. Examples of AMM models are the so-called 
Constant Function Market Makers (CFMMs) [1],[2],[9], which employ a suitable invariant mapping as the trading function.
The well-known Uniswap [3] is in turn an example of a CFMM AMM for which a constant-product formula is used to define
valid transactions for the model. Each trade must be executed in such a way that the quantity removed with respect to one
asset in a trade is compensated for by the quantity of the other asset added. Moreover, a trading function defined by means of a 
(weighted) geometric mean gives rise to a CFMM [7] closely related to the constant product CFMM. Alternately, instead of exploiting a constant product 
or geometric mean trading function, constant sum or arithmetic mean trading functions can also be used to define a CFMM, 
resulting in a constant sum or arithmetic mean CFMM [6]. 

In this article, we describe new forms of CFMMs we call the Generalized Mean Market Makers (G3Ms). We note that the G3Ms were presented (by the author of this article) in the 2021 US patent filing [5]. 
The trading functions of these G3Ms will respectively
correspond to what are known in mathematics as the Generalized Means [4] -- a parametrized family of averages that extends and generalizes
the conventional geometric mean as well as the standard arithmetic mean. Let $x_{1},...,x_{n}$ be $n$ given nonnegative real 
numbers, and assume that $n$ nonnegative weights $w_{1}, ...,w_{n}$ satisfying
\begin{equation}
w_{1}+...+w_{n}=1
\end{equation}
are also given. Then define the Generalized Mean with respect to a given real number $p\neq 0$ via 
\begin{equation}
\mu_{p}(\mbox{\boldmath $x$})=\mu_{p, \mbox{\boldmath $w$}}(\mbox{\boldmath $x$})=\left(\sum_{i=1}^{n}w_{i}x_{i}^{p}\right)^{\frac{1}{p}}.
\end{equation}
where $\mbox{\boldmath $w$}=(w_{1},...,w_{m})$ and $\mbox{\boldmath $x$}=(x_{1},...,x_{n})$. Moreover, at $p=0$, define
\begin{equation}
\mu_{0}(\mbox{\boldmath $x$})=\mu_{0,\mbox{\boldmath $w$}}(\mbox{\boldmath $x$})=\Pi^{n}_{i=1}x_{i}^{w_{i}}.
\end{equation}
Hence the Generalized Mean $\mu_{0}(\mbox{\boldmath $x$})$ at $p=0$ coincides with the (weighted) geometric mean, and, at $p=1$, 
the Generalized Mean $\mu_{1}(\mbox{\boldmath $x$})$ coincides with the (weighted) arithmetic mean. 
We also extend, in this paper, the concept of a Generalized Mean Market Maker by introducing the Generalized f-Mean Market Makers (Gf3Ms) as we explain below.

We expect G3Ms for the intermediate values of $p$ with $0<p<1,$ for example, to exhibit properties intermediate between the geometric and
arithmetic means and hence to potentially exhibit more favorable behavior as AMMs in at least some cases than either of the 
two end point models independently can. In fact, this intuition appears to be strongly supported by the following

\noindent\bf Key Fact.\rm We have
\begin{equation}
\mbox{\rm lim}_{p\rightarrow 0}\mu_{p, \mbox{\boldmath $w$}}(\mbox{\boldmath $x$})=\mu_{0, \mbox{\boldmath $w$}}(\mbox{\boldmath $x$}).
\end{equation}
Thus we see in (4) that, for any vector $\mbox{\boldmath $x$}$ and set of weights $\mbox{\boldmath $w$}$,  the corresponding generalized means at $p$ 
actually converge as $p\rightarrow 0$  to the associated generalized mean at $0$ (i.e., the corresponding geometric mean value). This is a well-known result, stated for 
example in [4] in the uniform-weight case (indeed the proof, which can be based on L'Hopital's rule, is standard so that we refrain from repeating it here).

A key metric (i.e., figure of merit) to consider when assessing the effectiveness of different AMM (CFMM) models is the slippage, 
which may be defined as the difference between the expected cost of an order to trade a given asset and the cost 
actually incurred at the time the order executes (see \S 3 below for a formal definition). Generally speaking, a slippage relatively low in absolute value is 
considered more favorable. As is well-known (again see, for example, our own \S 3), the Arithmetic Mean CFMM 
is easily shown to exhibit $0$ slippage in principle, but, by virtue of the way it is defined mathematically, can only
support trades whose total cost is bounded above by a fixed value -- a serious limitation. The geometric mean CFMM, on the other hand, can
in fact support trades of arbitrarily high value but unfortunately features non-zero slippage.
In this article, we also present and prove Theorem 1, in \S 3. This theorem demonstrates the significance of extending the 
Arithmetic Mean and Geometric Mean CFMMs as we do in this paper to intermediate G3M models with $p$-values in the range $0<p<1.$ It demonstrates this
by showing that, somewhat loosely speaking, there exists a sequence of G3M models 
for certain values of $p,0<p<1,$ such that, respectively, valid (buy) trades of arbitrarily large size can be supported but for 
which the respective slippage values for these models grow in absolute value significantly more slowly than the corresponding slippage values 
for the Geometric Mean CFMM model (that is, for the G3M model at $p=0$).  Thus, the G3M model
for intermediate values of $p,0<p<1,$ offers a significant advantage over the G3M model at $p=1$ because trades of aribtrarily large size can be supported, 
yet, at the same, also offers an advantage over the G3M model at $p=0$ as well due to significantly slower slippage growth.

The rest of the paper is organized as follows. In \S 2.1 we formally introduce the Generalized Mean Market Maker (G3M) models and prove that they satisfy
a concavity property (see Lemma 1) that valid CFMMs are expected to possess [1],[9]. In \S 2.2, we introduce the Generalized f-Mean Market Makers (Gf3Ms). Finally, in \S 3, 
we formally present and prove Theorem 1.

\section{Generalized Mean and Generalized f-Mean\\ Market Makers}
\subsection{Generalized Mean Market Makers}
In this subsection we formally present the parametrized family of Constant Function Market Makers (CFMMs) we call the Generalized Mean Market Makers (G3Ms). Our exposition here in \S 2 as well as in the rest of the paper -- in terms of the basic underlying definitions and  concepts -- is based on those in [1], [2], [9].

To define G3Ms,  we will first need to introduce the fundamental concept of a trading function of a CFMM. We begin, however, with a few requisite definitions. For any positive integer $n$, consider a basket of $n$ cryptocurrencies, intended for trading, to which is associated a vector  $\mbox{\boldmath $\cal{R}$}=(R_{1},...,R_{n})\in (\mbox{\boldmath $R$}_{+}\cup\{0\})^{n}$, where, here,  $\mbox{\boldmath $R$}_{+}$ denotes the set of positive real numbers. $\mbox{\boldmath $\cal{R}$}$ is called the (cryptocurrency) reserves, and each quantity $R_{i}$ is the reserve amount -- which we think of as being a fixed value -- in the exchange of currency $i,i=1,...,n,$ respectively. We denote by $\mbox{\boldmath $\Delta$}=(\Delta_{1},...,\Delta_{n})\in (\mbox{\boldmath $R$}_{+}\cup\{0\})^{n}$ the (cryptocurrency) input trade, so that $\Delta_{i}$ is the amount of currency $i$ that a trader or market participant proposes to tender or offer to the DEX in exchange for another currency or currencies. Furthermore, denote by $\mbox{\boldmath $\Lambda$}=(\Lambda_{1},...,\Lambda_{n})\in (\mbox{\boldmath $R$}_{+}\cup\{0\})^{n}$ the output trade, with  $\Lambda_{i}$ being the respective amount of the output trade in currency $i,i=1,...,n,$ so that $\Lambda_{i}$ will be the amount of currency $i$ that is proposed to be received by the trader from the DEX in return should the trade be executed. We call $(\mbox{\boldmath $\Delta$},\mbox{\boldmath $\Lambda$})$ a trade or proposed trade.

A trading function, denoted by $\tau$, is defined by 
\begin{eqnarray}
\tau:(\mbox{\boldmath $R$}_{+}\cup\{0\})^{n}\times(\mbox{\boldmath $R$}_{+}\cup\{0\})^{n}\times(\mbox{\boldmath $R$}_{+}\cup\{0\})^{n}\rightarrow\mbox{\boldmath $R$}\\
\tau(\mbox{\boldmath $\cal{R}$},\mbox{\boldmath $\Delta$},\mbox{\boldmath $\Lambda$})=\mu(R_{1}+\Delta_{1}-\Lambda_{1},...,R_{n}+\Delta_{n}-\Lambda_{n}),
\end{eqnarray}
for some given function $\mu:(\mbox{\boldmath $R$}_{+}\cup\{0\})^{n}\rightarrow \mbox{\boldmath $R$}$ (whose properties we will have more to say about below in this section). 

The trading function $\tau$ specifies whether a trade is regarded as legitimate and hence may be executed. Indeed, in the CFMM setting a proposed trade $(\mbox{\boldmath $\Delta$},\mbox{\boldmath $\Lambda$})$ is legitimate and may be executed if
it satisfies
\begin{equation}
\tau(\mbox{\boldmath $\cal{R}$},\mbox{\boldmath $\Delta$},\mbox{\boldmath $\Lambda$})=\tau(\mbox{\boldmath $\cal{R}$},\mbox{\boldmath $0$},\mbox{\boldmath $0$})=C. 
\end{equation}
where $C>0$. In other words, the trade is accepted only if the trading function is maintained at the constant value $\tau(\mbox{\boldmath $\cal{R}$},\mbox{\boldmath $0$},\mbox{\boldmath $0$})=C$. Indeed we can regard a CFMM as defined by its trading function along with its reserves $\mbox{\boldmath $\cal{R}$}$.

A standard example within the AMM/CFMM context here of a trading function as given by (5)-(6)  is the geometric mean CFMM trading function
\begin{equation}
\tau(\mbox{\boldmath $\cal{R}$},\mbox{\boldmath $\Delta$},\mbox{\boldmath $\Lambda$})=\sqrt[n]{\Pi^{n}_{i=1}(R_{i}+\Delta_{i}-\Lambda_{i})},
\end{equation}
so that with respect to (8) the definition of the function $\tau$ is based on that of the geometric mean as given by  (3) with $w_{i}=1/n, i=1,...,n$. Another such example is the arithmetic mean CFMM defined by
\begin{equation}
\tau(\mbox{\boldmath $\cal{R}$},\mbox{\boldmath $\Delta$},\mbox{\boldmath $\Lambda$})=\frac{1}{n}\sum^{n}_{i=1}R_{i}+\Delta_{i}-\Lambda_{i},
\end{equation} 
so that with respect to (9) the definition of the function $\tau$ is now based on that of the arithmetic  mean as given by  (2) with $p=1$ and $w_{i}=1/n,i=1,...,n.$

We now formally introduce the family of CFMMs we call the Generalized Mean Market Makers (G3Ms) by defining their respective trading functions $\tau=\tau_{p}$ at each $p,0<p\leq 1$, in the following way:
\begin{equation}
\tau_{p}(\mbox{\boldmath $\cal{R}$},\mbox{\boldmath $\Delta$},\mbox{\boldmath $\Lambda$})=\left(\sum_{i=1}^{n}w_{i}(R_{i}+\Delta_{i}-\Lambda_{i})^{p}\right)^{\frac{1}{p}}.
\end{equation}
for any given, fixed set of weights $w_{1},...,w_{n}$, so that $\mu=\mu_{p}$, where the function $\mu$ is as in (6) and $\mu_{p}$ is as in (2). In addition, the G3M trading function $\tau=\tau_{0}$ at $p=0$ is defined by 
\begin{equation}
\tau_{0}(\mbox{\boldmath $\cal{R}$},\mbox{\boldmath $\Delta$},\mbox{\boldmath $\Lambda$})=\Pi_{i=1}^{n}(R_{i}+\Delta_{i}-\Lambda_{i})^{w_{i}}
\end{equation}
again for any given, fixed set of weights $w_{1},...,w_{n}$, so that, in this case in (11), $\mu=\mu_{0}$, where $\mu_{0}$ is as in (3).

Hence within the G3M model, for any $p,0<p\leq 1$, equations (7) become
\begin{equation}
\left(\sum_{i=1}^{n}w_{i}(R_{i}+\Delta_{i}-\Lambda_{i})^{p}\right)^{\frac{1}{p}}=\left(\sum_{i=1}^{n}w_{i}R_{i}^{p}\right)^{\frac{1}{p}}=C,
\end{equation}
and, at $p=0,$
\begin{equation}
\Pi_{i=1}^{n}(R_{i}+\Delta_{i}-\Lambda_{i})^{w_{i}}=\Pi_{i=1}^{n}R_{i}^{w_{i}}=C.
\end{equation}

According to [1] and [9], the properties of being concave, nondecreasing, nonnegative, and differentiable characterize the (multivariate) function $\mu$ in (6) as defining a suitable CFMM trading function $\tau$.  Now it is clear that the function $\mu=\mu_{p, \mbox{\boldmath $w$}}$ as in (2)-(3), for a fixed set of weights $\mbox{\boldmath $w$}=(w_{1},...,w_{n})$, and any given $p,0\leq p\leq 1$ is in fact nonnegative and nondecreasing on $(\mbox{\boldmath $R$}_{+}\cup\{0\})^{n}$ (given any reasonable extension to several variables of the definition of a nondecreaing function, such as that the function is nondecreasing in any one variable whenever all of the others are held constant). Next, by a standard result $\mu_{p, \mbox{\boldmath $w$}}$ is also differentiable on $\mbox{\boldmath $R$}_{+}^{n}$ since all of its (first-order) partial derivatives exist and are continuous. 

As for concavity of the function $\mu=\mu_{p, \mbox{\boldmath $w$}}$ for any fixed set of weights $\mbox{\boldmath $w$}=(w_{1},...,w_{n})$, let's first recall the definition of a concave function. A function $f:D\rightarrow \mbox{\boldmath $R$}$, defined on the convex set $D\subseteq\mbox{\boldmath $R$}^{n}$ is said to be a concave function if, for all 
$t\in [0,1]$ and $\mbox{\boldmath $x$},\mbox{\boldmath $y$}\in D$, 
\begin{equation}
f((1-t)\mbox{\boldmath $x$}+t\mbox{\boldmath $y$})\geq (1-t)f(\mbox{\boldmath $x$})+tf(\mbox{\boldmath $y$}).
\end{equation}
The following lemma establishes concavity of the functions $\mu_{p}=\mu_{p, \mbox{\boldmath $w$}},0\leq p\leq 1.$

\noindent\bf Lemma 1.\rm\it\,For each $p,0\leq p\leq 1,$ the function $\mu_{p}=\mu_{p,\mbox{\boldmath $w$}}:(\mbox{\boldmath $R$}_{+}\cup \{ 0\})^{n}\rightarrow \mbox{\boldmath $R$}$ as in (2)-(3) is a concave function for any fixed set of weights $\mbox{\boldmath $w$}=(w_{1},...,w_{n})$ as in (1).\rm

\noindent\rm\bf Proof.\,\rm First let's show this for all $p,0<p\leq 1.$ Define, for any $\mbox{\boldmath $x$}=(x_{1},...,x_{n})\in (\mbox{\boldmath $R$}_{+}\cup \{ 0\})^{n}$
\begin{equation}
\overline{\mu}_{p}(\mbox{\boldmath $x$})=\left(\sum_{i=1}^{n}x_{i}^{p}\right)^{\frac{1}{p}}.
\end{equation}
Since
\begin{equation}
\left(\sum_{i=1}^{n}w_{i}x_{i}^{p}\right)^{\frac{1}{p}}=\left(\sum_{i=1}^{n} (w_{i}^{1/p}x_{i})^{p}\right)^{\frac{1}{p}}
\end{equation}
and
\begin{equation}
\overline{\mu}_{p}(t\mbox{\boldmath $x$})=t\overline{\mu}_{p}(\mbox{\boldmath $x$}),\mbox{\rm for any}\,t\geq 0,
\end{equation}
it follows that, in order to establish the lemma, we need only prove that
\begin{equation}
\overline{\mu}_{p}(\mbox{\boldmath $x$})+\overline{\mu}_{p}(\mbox{\boldmath $y$})\leq\overline{\mu}_{p}(\mbox{\boldmath $x$}+\mbox{\boldmath $y$})
\end{equation}
for any $\mbox{\boldmath $x$},\mbox{\boldmath $y$}\in (\mbox{\boldmath $R$}_{+}\cup \{ 0\})^{n}$, where $\mbox{\boldmath $x$}=(x_{1},...,x_{n})$ and
$\mbox{\boldmath $y$}=(y_{1},...,y_{n}).$ Let $a_{i}=x_{i}^{p},$ and $b_{i}=y_{i}^{p}$ and raise both sides of (18) to the $p$ power to obtain
\begin{equation}
\left(\left(\sum^{n}_{i=1}a_{i}\right)^{1/p}+\left(\sum^{n}_{i=1}b_{i}\right)^{1/p}\right)^{p}\leq \sum^{n}_{i=1}(a^{1/p}_{i}+b^{1/p}_{i})^{p}.
\end{equation}
But, (19) is equivalent to the following:
\begin{equation}
||(\sum^{n}_{i=1}a_{i},\sum^{n}_{i=1}b_{i})||_{1/p}=||\sum^{n}_{i=1}(a_{i},b_{i})||_{1/p}\leq \sum^{n}_{i=1}||(a_{i},b_{i})||_{1/p}.
\end{equation}
Now note that (20) is clearly true applying the triangle inequality for the $(1/p)$-norm in order to obtain the inequality in (20), so the lemma for all $p,0<p\leq 1,$ follows.

Finally we extend our concavity proof for the functions  $\mu_{p},0<p\leq 1,$ to the function $\mu_{0}$. For this, assume that $\mu_{0}$ is not concave. 
Then there exist vectors $\mbox{\boldmath $w$},\mbox{\boldmath $x$}, \mbox{\boldmath $y$}\in\mbox{\boldmath $R$}^{n}$ as well as $t\in [0,1]$ for which 
\begin{equation}
\mu_{0}((1-t)\mbox{\boldmath $x$}+t\mbox{\boldmath $y$})< (1-t)\mu_{0}(\mbox{\boldmath $x$})+t\mu_{0}(\mbox{\boldmath $y$}).
\end{equation}
But by the Key Fact (4) we could choose a $p,0<p\leq 1,$ small enough that (21) would also imply that 
\begin{equation}
\mu_{p}((1-t)\mbox{\boldmath $x$}+t\mbox{\boldmath $y$})< (1-t)\mu_{p}(\mbox{\boldmath $x$})+t\mu_{p}(\mbox{\boldmath $y$}).
\end{equation}
But this violates the concavity of $\mu_{p}$ that we just established above, completing our proof of the lemma in its entirety.\,\,$\Box$

Additionally, according to [1], homogeneity is also a common property of the function $\mu$ defining the trading function.
In fact, we call a function $f:(\mbox{\boldmath $R$}_{+}\cup \{ 0\})^{n}\rightarrow \mbox{\boldmath $R$}$ homogeneous (of first order) if it satisfies 
\begin{equation}
f(tx_{1},...,tx_{n})=tf(x_{1},...,x_{n}),\,\,\mbox{\rm for any}\,\,\mbox{\boldmath $x$}=(x_{1},...,x_{n})\in (\mbox{\boldmath $R$}_{+}\cup \{ 0\})^{n},\, t\in \mbox{\boldmath $R$}_{+}.
\end{equation}
It is clear from the definition (23) that each function $\mu_{p, \mbox{\boldmath $w$}},0\leq p\leq 1$, is homogeneous of first order as well.
\subsection{Generalized f-Mean Market Makers (Gf3Ms)}
We can even generalize the G3M model as described above to the case of CFMMs characterized by trading functions defined by means of a class of functions that extends the Generalized Means.
This class of functions extending the Generalized Means is the so-called set of (weighted) Generalized f-Means (GfMs) (see for example [8]), which are defined by
\begin{displaymath}
M_{f,\mbox{\boldmath $w$}}(\mbox{\boldmath $x$})=f^{-1}(w_{1}f(x_{1})+...+w_{n}f(x_{n}))
\end{displaymath}
for $\mbox{\boldmath $x$}=(x_{1},...,x_{n})\in (\mbox{\boldmath $R$}_{+}\cup\{0\})^{n}$ and $\mbox{\boldmath $w$}=(w_{1},...,w_{m})\in (\mbox{\boldmath $R$}_{+}\cup\{0\})^{n}$ (with (1) holding as before),  as well as a chosen continuous and injective function $f$ mapping some interval $I\subseteq\mbox{\boldmath $R$}$ into $\mbox{\boldmath $R$}$. Here, in the definition of $M_{f,\mbox{\boldmath $w$}}$ above, ``$f^{-1}$" refers to the inverse function with respect to $f$, so that $f\circ f^{-1}=f^{-1}\circ f=$Id. The GfMs give rise to a corresponding class of CFMMs which we call the Generalized f-Mean Market Makers (Gf3Ms), in particular if $M_{f,\mbox{\boldmath $w$}}$ satisfies the conditions of being concave, nondecreasing, nonnegative, and differentiable as discussed in \S 2.1.

The Generalized Means are special cases of the Generalized f-Means, and hence the G3Ms are special cases of Gf3Ms. Indeed the Generalized Mean at each $p\in\mbox{\boldmath $R$},p\neq 0$, clearly coincides with the GfM for $f\equiv x^{p}$, respectively. It is moreover easy to see that the Generalized Mean at $p=0$ (that is, the geometric mean) coincides with the GfM with $f\equiv \mbox{\rm log}_{e}(x)$. We note that the Generalized f-Mean is also called the Quasi-Arithmetic Mean as well as the Kolmogorov Mean in the literature. 

\section{Balancing Slippage and Trade Size for G3Ms}
In this section, we work within the notational and conceptual framework of \S 2 but specialize to the case of G3Ms for which $n=2, \Lambda_{1}=\Delta_{2}=0$, and $w_{1}=w_{2}=1/2$. We show -- see in particular our Theorem 1 as well as Remark 1 below -- how our G3M models, by taking on intermediate values of $p$ for which $0<p<1,$ can balance slippage (the formal definition of which follows below) against admissible trade size in such a way that these models can be superior to both the geometric mean (constant product) and the arithmetic mean (constant sum) CFMMs with respect to both of these two metrics (i.e., slippage and admissible trade size) simultaneously. Indeed, we see from (12) that
\begin{equation}
\Delta_{1}=\left(2C^{p}-(R_{2}-\Lambda_{2})^{p}\right)^{1/p} -R_{1}
\end{equation}
Since $C, R_{1},$ and $R_{2}$ are considered fixed in the CFMM framework and $0\leq \Lambda_{2}\leq R_{2}$ as well, we see in turn from (24) that, for all $p,0<p\leq 1,$ the G3M model at $p$ can only support an input trade $\Delta_{1}$ of bounded size, seriously limiting liquidity of the exchange. On the other hand, it is clear from the nature of the product operation that the geometric mean CFMM as in (11) or (13) can support such trades of arbitrary size (i.e., arbitrarily large value or price), offering a key advantage. However, as we shall see in Theorem 1 below (and in any case one might already expect given (4)), the G3M models with $0<p<1$ can also support trades of arbitrarily large size as $p\rightarrow 0$.

So, in the setting of this section within which we presuppose that $n=2, \Lambda_{1}=\Delta_{2}=0$, and $w_{1}=w_{2}=1/2$ as stated above, (10) becomes 
\begin{equation}
\tau_{p}(\mbox{\boldmath $\cal{R}$},\Delta_{1},\Lambda_{2})=\left(\frac{(R_{1}+\Delta_{1})^{p}}{2}+\frac{(R_{2}-\Lambda_{2})^{p}}{2}\right)^{1/p},
\end{equation}
for $p,0<p\leq 1, $ and  (12) becomes
\begin{equation}
\left(\frac{(R_{1}+\Delta_{1})^{p}}{2}+\frac{(R_{2}-\Lambda_{2})^{p}}{2}\right)^{1/p}=\left(\frac{R_{1}^{p}}{2}+\frac{R_{2}^{p}}{2}\right)^{1/p}=C.
\end{equation}
At $p=0,$ we have 
\begin{equation}
\tau_{0}(\mbox{\boldmath $\cal{R}$},\Delta_{1},\Lambda_{2})=\sqrt{(R_{1}+\Delta_{1})(R_{2}-\Lambda_{2})},
\end{equation}
and (13) becomes
\begin{equation}
\sqrt{(R_{1}+\Delta_{1})(R_{2}-\Lambda_{2})}=\sqrt{R_{1}R_{2}}=C.
\end{equation}

Now, following [9], define the spot exchange rate with respect to the trading function $\tau=\tau_{p}, 0\leq p\leq 1,$ via 
\begin{equation}
E_{p}=E_{p,12}=E_{p,12}(\mbox{\boldmath $\cal{R}$},0,0)=\frac{\frac{\partial\tau_{p}}{\partial R_{2}}(\mbox{\boldmath $\cal{R}$},0,0)}{\frac{\partial\tau_{p}}{\partial R_{1}}(\mbox{\boldmath $\cal{R}$},0,0)}.
\end{equation}
We then define, again following [9], the slippage -- with respect to the variables $\Delta_{1}$ and $\Lambda_{2}$ -- as
\begin{equation}
S_{p}=S_{p,12}=\frac{\Delta_{1}}{\Lambda_{2}}E_{p}^{-1} -1,
\end{equation}
for $p,0\leq p\leq 1$ (provided of course $|E_{p}|>0$). So, at $p=0$, we can calculate that
\begin{equation}
S_{0}=\frac{R_{2}}{R_{1}}\frac{\Delta_{1}}{\Lambda_{2}}-1.
\end{equation}
Hence, considering (28) as well,
\begin{equation}
S_{0}=\frac{\left(\frac{C^{2}}{(R_{2}-\Lambda_{2})}-R_{1}\right)}{\Lambda_{2}}\frac{R_{2}}{R_{1}}-1
\end{equation}
For $p,0<p\leq 1,$ we have
\begin{eqnarray}
S_{p}&=&\frac{R_{1}^{p-1}}{R_{2}^{p-1}}\frac{\Delta_{1}}{\Lambda_{2}}-1\nonumber\\
&=& \frac{R_{1}^{p-1}}{R_{2}^{p-1}}\frac{(2C^{p}-(R_{2}-\Lambda_{2})^{p})^{1/p}-R_{1}}{\Lambda_{2}}-1
\end{eqnarray}

In the rest of the paper, we will often write
\begin{equation}
\epsilon=R_{2}-\Lambda_{2}.
\end{equation} 
That is,
\begin{equation}
\Lambda_{2}=\Lambda_{2}(\epsilon)=R_{2}-\epsilon>0.
\end{equation} 
for any $\epsilon,0<\epsilon< R_{2}$. We think of the most interesting cases in this context to be the ones for which $\epsilon>0$ is small, which enables, by (24), the corresponding quantity $R_{1}+\Delta_{1}=R_{1}+\Delta_{1}(\epsilon)$ and hence also the quantity $\Delta_{1}=\Delta_{1}(\epsilon)$ to be large (or at least relatively large), where $\Delta_{1}=\Delta_{1}(\epsilon)$ satisfies
\begin{equation}
\left(\frac{(R_{1}+\Delta_{1}(\epsilon))^{p}}{2}+\frac{\epsilon^{p}}{2}\right)^{1/p}=C
\end{equation}
for given $p,0<p\leq 1$. That is, for this range of $p$,
\begin{equation}
\Delta_{1}(\epsilon)=(2C^{p}-\epsilon^{p})^{1/p}-R_{1}.
\end{equation}

So, with these definitions we can rewrite (32) and (33) in the form:

\begin{eqnarray}
S_{0}=S_{0}(\epsilon)&=&\frac{\left(\frac{C^{2}}{\epsilon}-R_{1}\right)}{\Lambda_{2}(\epsilon)}\frac{R_{2}}{R_{1}}-1\\
S_{p}=S_{p}(\epsilon)&=&\frac{R_{1}^{p-1}}{R_{2}^{p-1}}\frac{(2C^{p}-\epsilon^{p})^{1/p}-R_{1}}{\Lambda_{2}(\epsilon)}-1.
\end{eqnarray}
Note that, at $p=1$, we must have $\Delta_{1}=\Lambda_{2}$, so it is easy to see from (39) that $S_{1}=0$.

Before stating Theorem 1, we recall that a function $f:\mbox{\boldmath $R$}_{+}\rightarrow \mbox{\boldmath $R$}$ is $f(t)=O(g(t))$ for $t\rightarrow 0$ for some function 
$g:\mbox{\boldmath $R$}_{+}\rightarrow \mbox{\boldmath $R$}_{+}$ if there exist numbers $k>0$ and $t_{0}>0$ 
for which $|f(t)|\leq kg(t)$ for all $t,0<t\leq t_{0}$. Furthermore, we say that a function $f:\mbox{\boldmath $R$}_{+}\rightarrow \mbox{\boldmath $R$}$ is $f(t)=\Omega(g(t))$ for $t\rightarrow 0$ for some function 
$g:\mbox{\boldmath $R$}_{+}\rightarrow \mbox{\boldmath $R$}_{+}$ if there exist numbers $k>0$ and $t_{0}>0$ 
for which $|f(t)|\geq kg(t)$ for all $t,0<t\leq t_{0}$.

\noindent\bf Theorem 1.\rm\it\,Suppose that $C$ is such that $2<C<\infty$, and, define, for any desired $s,1<s<\frac{C}{2}$, the function $p:(0,1)\rightarrow\mbox{\boldmath $R$}_{+}$ via
\begin{equation}
 p(\epsilon)=p_{s}(\epsilon)=\frac{\mbox{\rm log}(s+\sqrt{s^{2}-s})}{\mbox{\rm log}\left(\frac{C}{\epsilon}\right)} \in (0,1]\,\,\,\mbox{\rm for}\,\,\,0<\epsilon<1.
\end{equation}
Then, we have, with the slippage $S_{p}(\epsilon)=S_{p(\epsilon)}(\epsilon)$ at $p=p(\epsilon)$ defined as in (39),
\begin{equation}
|S_{p(\epsilon)}(\epsilon)|=O(\epsilon^{-c})
\end{equation}
as $\epsilon\rightarrow 0$, where $c$ satisfies $0<c<1$ and we can in fact take
\begin{equation}
c=c(s)=1-\frac{\mbox{\rm log}(s)}{\mbox{\rm log}(s+\sqrt{s^{2}-s})}.
\end{equation}
Moreover, defining $\Delta_{1}=\Delta_{1}(\epsilon)$ as in (37) with $p=p(\epsilon)$ there defined once again as in (40), we have
 \begin{equation}
\Delta_{1}(\epsilon)=\Omega(\epsilon^{-c})
\end{equation}
as  $\epsilon\rightarrow 0$, where, in (43), c=c(s) is defined once again as in (42) above.\rm 

\noindent\rm\bf Remark 1.\,\rm  Before we move on to the proof of Theorem 1, we note that, with slippage $S_{0}(\epsilon)$ at $p=0$ defined as in (38), we have from (38) that
\begin{equation}
|S_{0}(\epsilon)|=\Omega(\epsilon^{-1}).
\end{equation}
as $\epsilon\rightarrow 0$. Hence, based on (41), the intermediate G3M models with $p=p_{s}(\epsilon)$,  for any $s, 1<s<\frac{C}{2},$ are seen to offer superior slippage properties in this case as $\epsilon\rightarrow 0$ relative to the G3M model in this setting at $p=0$ (that is, the Constant Product/Geometric Mean CFMM). Meanwhile, (43) implies that, furthermore, these intermediate G3M models with $p=p_{s}(\epsilon)$ as $\epsilon\rightarrow 0$ support aribitrarily large trade sizes also as $\epsilon\rightarrow 0$, making them superior as well in this respect to the G3M model at $p=1$ (that is, the Constant Sum/Arithmetic Mean CFMM).

\noindent\rm\bf Proof of Theorem 1.\,\rm  Given $s,1<s<\frac{C}{2},$ as in the theorem's statement, to establish the theorem we will determine $p,0<p\leq 1,$ for which
\begin{equation}
(2C^{p}-\epsilon^{p})^{1/p}=\frac{C^{2}}{s^{1/p}\epsilon}.
\end{equation}
Hence we want to solve
\begin{equation}
2C^{p}-\epsilon^{p}=\frac{C^{2p}}{s\epsilon^{p}}.
\end{equation}
That is, we want to solve
\begin{equation}
x^{2}-2sx+s=0,
\end{equation}
where $x=\frac{C^{p}}{\epsilon^{p}}$, So, solving the quadratic equation (47) for $x$, we obtain
\begin{equation}
x=s\pm \sqrt{s^{2}-s}.
\end{equation}
Hence if we choose $p>0$ such that
\begin{equation}
\frac{C^{p}}{\epsilon^{p}}=x=s +\sqrt{s^{2}-s},
\end{equation}
then (46) will be satisfied. So, we can take $p=p(\epsilon)$ to be
\begin{equation}
p=p(\epsilon)=\frac{\mbox{\rm log}(x)}{\mbox{\rm log}(\frac{C}{\epsilon})}=\frac{\mbox{\rm log}(s+\sqrt{s^{2}-s})}{\mbox{\rm log}(\frac{C}{\epsilon})}
\end{equation}
and then (45) will be satisfied for $p=p(\epsilon)$ as in (50). So, from (45), we get
\begin{eqnarray}
(2C^{p}-\epsilon^{p})^{1/p}&=&\frac{C^{2}}{s^{1/p}\epsilon}\nonumber\\
&=& \frac{C^{2}}{\left(s^{\frac{\mbox{\rm log}(\frac{C}{\epsilon})}{\mbox{\rm log}(s+\sqrt{s^{2}-s})}}\right)\epsilon}\nonumber\\
&=& \frac{C^{2}}{\left((\frac{C}{\epsilon})^{\frac{\mbox{\rm log}(s)}{\mbox{\rm log}(s+\sqrt{s^{2}-s})}}\right)\epsilon^{\frac{\mbox{\rm log}(s+\sqrt{s^{2}-s})}{\mbox{\rm log}(s+\sqrt{s^{2}-s})}}}\nonumber\\
&=& \frac{C^{(2-\mbox{log}(s))}}{\epsilon^{\left(1-\frac{\mbox{\rm log}(s)}{\mbox{\rm log}(s+\sqrt{s^{2}-s})}\right)}},
\end{eqnarray}
for any desired $s$ as in the statement of the theorem. This establishes (43). Finally, (41) follows directly from (51) as well since, writing $R_{2}/R_{1}=R$, it is the case that, for fixed $s$,
\begin{equation}
R^{p(\epsilon)-1}=O(1).
\end{equation}
as $\epsilon\rightarrow 0$.\,\,$\Box$

\begin{center}
\large\bf References
\end{center}
\noindent [1] Angeris, G. Agrawal, A., Evans, A., Chitra, T., and Boyd, S. (2021) Constant function market makers: Multi-asset trades via convex optimization.
arXiv preprint arXiv:2107.12484.

\noindent [2] Angeris, G. and Chitra, T. (2020) Improved price oracles: Constant function market makers. In it\,Proceedings
of the 2nd ACM Conference on Advances in Financial
Technologies,\rm 80–91.

\noindent [3] Angeris, G., Kao, H.T., Chiang, R., Noyes, C., and Chitra, T. (2019) An analysis of Uniswap
markets. arXiv preprint arXiv:1911.03380.

\noindent [4] Berger, R. and Casella, G. (1992) Deriving Generalized Means as Least Squares and Maximum Likelihood Estimates \it The American Statistician,\rm
Vol. 46, No. 4, pp. 279-282.

\noindent [5] Karlin, M., Zanger, D., and Katz, A. (2021) (US patent filing) U.S. App. No. 17/818,847, “Systems and Methods for Automated Staking Models” as originally-filed as U.S. Provisional App. No. 63/283,885 (November 29, 2021).

\noindent [6] Krishnamachari, B., Feng, Q., and Grippo, E. (2021) Dynamic Curves for Decentralized Autonomous Cryptocurrency Exchanges. arXiv preprint
arXiv:2101.02778 

\noindent [7] Martinelli, F. and Mushegian, N. (2019) Balancer: A non-custodial portfolio manager, liquidity provider, and price sensor.

\noindent [8] Matkowski, J. and Pales, Z. (2015) Characterization of generalized quasi-arithmetic means. Acta 
\it Scientiarum Mathematicarum\rm, 81(3-4):447–456.

\noindent [9] Xu, J., Paruch, K., Cousaert, S., and Feng, Y. (2021)  “SoK: Decentralized
Exchanges (DEX) with Automated Market Maker (AMM) protocols.” http://arxiv.org/abs/2103.12732

\end{document}